# COVID-19 DYNAMIC MODEL: BALANCED IDENTIFICATION OF GENERAL BIOLOGICAL AND COUNTRY SPECIFIC SOCIAL FEATURES

A.V.Sokolov[1], L.A.Sokolova[2]

[1] *Institute for information transmission problem (Kharkevitch Insitute) RAS,*

*Bolshoy Karetny per. 19, build.1, Moscow 127051 Russia*

*e-mail: alexander.v.sokolov@gmail.com*

[2] *Federal Research Center "Computer Science and Control" of RAS Institute for Systems*

*Analysis, pr. 60-letiya Oktyabrya 9, Moscow 117312 Russia*

*e-mail: las.sokolova@gmail.com*

Breaking a complex bio-social phenomenon (epidemic) into its components, considering the processes that determine its dynamics, formalizing the accepted hypotheses in mathematical equations, selecting appropriate experimental and statistical material, and constructing a mathematical model – those are typical tasks of scientific research. A specific data processing method (balanced identification) and appropriate information technology made it possible to consider a number of models, determine the general biological laws of the virus-human interaction (common to all populations), and the country specific social features of epidemic management in the countries (or cities) under consideration. As the initial data, only new cases are used. Data from different countries is taken from official sources and processed in a uniform way. The obtained estimates of the number of undetected infected are lower estimates.

Keywords: mathematical modeling, COVID-19, balanced identification, bio-social system

## INTRODUCTION

The dynamics of epidemics (including COVID-19) is determined both by biological characteristics of the human body vs. virus interaction, and by social aspects of the interaction between man and society.

Biological features determine:
- "potential" number of people infected by one person,
- the probability of recovery (without isolation),
- "visibility" of symptoms (for identification and subsequent isolation).

Society can manage an epidemic in two ways:
- limit contacts (self-isolation) or make them safer (distancing, masks, etc.),
- identify and isolate those infected.

The paper uses population-based models of the virus spreading in human population (such as "host-parasite" or "predator-prey"). Particular attention is paid to the (clear) separation of the purely biological processes of interaction between the predator (virus) and the prey (human) and the purely social mechanisms of society counteracting an epidemic. When modeling the dynamics of the epidemic in different countries, biological functions are assumed to be the same and constant in time, while social functions can differ from country to country and vary in time.

The epidemic dynamics determined by undetected infected people (UDI) – those identified are isolated (more or less carefully) by the society, and those not identified continue to infect others. So, only the population of UDI is considered.

The typical time from infection to the moment when it is contagious (latent period) is about 4-5 days for COVID-19 (this estimate is not used in the study below), the typical modeling time is about 100 days. The characteristic times are comparable. Therefore, to describe the dynamics of UDI, it is necessary to divide the population into groups according to the time elapsed since the infection, which we will call the duration of infection (DI). Models of this type have long been known (see reviews by Svirezhev, Logofet, 1978; Hansen, 1986) and are widely used in demography, ecology and epidemiology.

A singe type of statistical information is used to build (identify) the models – the standard official data on the number of new (detected per day) cases of infection. Data is used "as is" without pre-processing. Countries with abnormal data (for example, China) are ignored. The number of countries present (7 populations) was determined by the limits of showing the simulation results in one figure clearly. Meanwhile the technology of balanced identification allows to increase the number several times. The constructed model considers 7 populations – the populations of Great Britain (Gbr), Germany (Deu), Italy (Ita), Spain (Esp), France (Fra), Russia excluding the city of Moscow and the Moscow Region (Rus-mos) and the city of Moscow including the Moscow Region (Mos). The division of Russia into two parts is caused by the significant difference in the dynamics.

The model under consideration was selected as the one that best fits the quantity and quality of the selected statistics. The balanced identification method used (Sokolov, Voloshinov, 2018; Sokolov, Voloshinov, 2019) allowed us to quantify how much the accepted set of hypotheses about the functioning of an object (bio-social system) corresponds to the available factual material (statistics).

## STATISTICAL INFORMATION USED

The model is based on statistics on new cases (of infection):
- UK, Germany, Italy and France – *https://ourworldindata.org/coronavirus-source-data*.
- Russia and Moscow – *https://www.rospotrebnadzor.ru/about/info/news_time/*
- Spain - *https://en.wikipedia.org/wiki/COVID-19_pandemic_in_Spain*

The statistics are normalized in time – for each country, the last date with the number of infected less than 100 was adopted as 0: for Great Britain analyses starts on 2020.03.05, Germany – 2020.02.29, Italy – 2020.02.23, Spain – 2020.03.02, France – 2020.02. 29, Russia (without Moscow and Moscow Region) and Moscow with Moscow Region – 2020.03.20.

Let's denote the entire array of statistics (dataset)

$$DS: \{ nC_{t,p}, \ t \in [0,T_p]; \ p \in [0,P], P=6 \},$$

where $nC_{t,p}$ is the number of new infections at time $t$ in a population (country or city) with number $p$.

The initial data for Moscow and the region are illustrated in Figure 1.

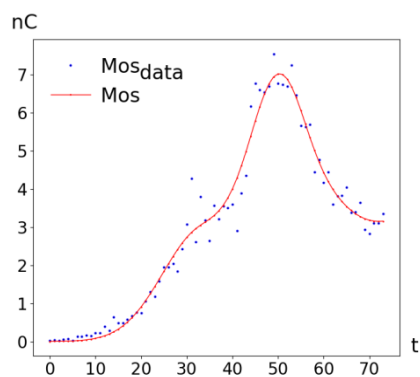

Figure 1. New cases of infection in Moscow and the region (thousand people) and the corresponding model curve: 2020.03.20 is taken as 0, the last point is 2020.05.30

# DYNAMICS MODEL OF A POPULATION OF UNDETECTED INFECTED PEOPLE WITH DISTRIBUTION BY DURATION OF INFECTION

This section is concerned with a difference model (such as the Leslie matrix; Svirezhev, Logofet, 1978; Hansen, 1986), its verbal description, and the discussion of accepted hypotheses. In parallel with the difference model in curly brackets without explanation, similar equations of continuous formulation are given. You can restrict yourself to a difference model or even just the verbal description of the accepted hypotheses. In addition, the Appendix provides a fragment of the formal model description in the task file, which is used to identify unknown model functions using balanced identification technology.

The model describes the dynamics of 7 UDP populations. For each population $p$, its own submodel is considered. The submodel is characterized by the functions $F_p(t)$ and $I_p(t)$, which describe social management. Common biological functions are $b(\tau)$, $\mu(\tau)$ and $f(\tau)$.

**Hypotheses.**

*The virus does not change. There are no mutations.* The functions ($b(\tau)$, $\mu(\tau)$, $f(\tau)$) that determine the interaction of the virus with the human body do not change over time.

*Immunity is not taken into account.* Only UDI are considered, the rest of the population is not taken into account in any way. While the percentage of those who have been ill is small, we can assume that the virus spreads in an endless population.

*Identified infected do not infect further.* Quarantine and isolation work: the number of those infected by identified infected (at the time of infection) is not large and can be neglected, taking into account the error of the initial data. Although among the medical staff there are a lot of cases.

*There is no migration.* Isolated populations are considered. It is assumed that infection occurs once at the initial moment and migration is not significant compared with the number of secondary infected. Apparently, the statement is questionable for the beginning of the modeling interval.

*The time step of the model.* A time step of 1 day corresponds to the statistical data update rate.

*Maximum DI.* $\tau_{max} = 16$ days. We assume that any UDI stops being contagious after 16 days. It corresponds to the quarantine time (2 weeks).

*Population.* A population of UDI is considered. The individuals of population can differ from each other at each time ($t$) by a single parameter – the time elapsed since the infection (DI), $\tau$. Sex and age structure, location, liability to disease, immunity etc. are not considered.

The number of UDI at time $t$ with a DI value of $\tau$ is denoted by $n(t, \tau)$.

The dynamics of such population consists of "DI parameter shift". For example, the number of people with a DI equal to 7 days on May 24 equals the number with a DI of 6 days on the previous day, May 23, minus recovered and identified (put into isolation, quarantined). Thus, for each of the 7 considered populations $p$,

$n_p(t+1, \tau+1) = n_p(t,\tau) - \mu(\tau) * n_p(t,\tau) - F_p(t) * f(\tau) * n_p(t,\tau)$

$$\left\{ \quad \frac{\partial n_p}{\partial t} + \frac{\partial n_p}{\partial \tau} = -\mu(\tau) n_p(t,\tau) - F_p(t) f(\tau) n_p(t,\tau) \quad \right\}$$

Figure 2 shows an example – the density of UDI for Moscow.

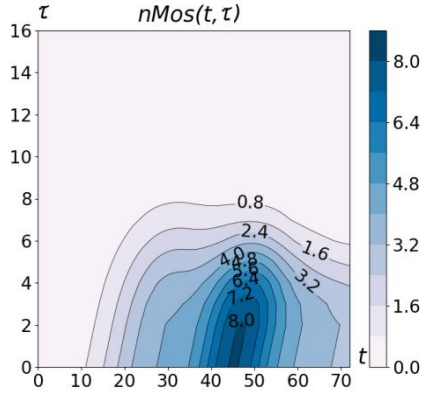

Figure 2. Density of UDI for Moscow and the region (lower estimate, thousand people) – $n(t, \tau)$ as of May 29

***Recovery.*** The recovery of UDI (and the cessation of the virus spread) is one of the causes for the decline in population. The number equals $\mu(\tau)*n_p(t,\tau)$, where $\mu(\tau) \leq 1$ is the probability of recovery for (DI) $\tau$. The first two days, the probability of recovery is 0:
$$\mu(0) = 0; \mu(1) = 0 \quad \{ \mu(\tau) = 0, \tau \in [0,2] \}.$$

***Case detection.*** Another reason for the decrease in the UDI population is case detection and isolation $F_p(t)*f(\tau) * n_p(t,\tau)$, where $F_p(t) * f(\tau) \leq 1$ is the detection probability, $f(\tau)$ is the biological factor (common for all countries), and $F_p(t)$ is social detection management – a factor that differs from country to country and varies over time.
The first two days, the probability of detection is 0:
 $f(0)=0; f(1)=0 \quad \{ f(\tau) = 0, \tau \in [0,2] \}.$
Every day, society efforts lead to a greater (or equal) detectability
 $F_p(t+1) \geq F_p(t) \quad \{ \frac{dF}{dt} \geq 0 \}.$
For definiteness, we normalize one of the factors:
$$\sum_{\tau'=0}^{16} f(\tau') = 1 \qquad \{ \int_0^{\tau_{max}} f(\tau') d\tau' = 1 \}$$

***Infectiousness (addition to the UDI population).*** The number of new UDI ("born" with zero duration of infection) is determined by the equation
$$n_p(t+1,0) = I_p(t) \sum_{\tau'=0}^{16} b(\tau') n_p(t,\tau') \qquad \{ n_p(t,0) = I_p(t) \int_0^{\tau_{max}} b(\tau') n_p(t,\tau') d\tau' \}$$
where $I_p(t) * b(\tau)$ is the number of new infected from one UDI with DI equal to $\tau$ (contagiousness or infectiousness of the undetected infected at time *t* with the duration of infection $\tau$), $b(\tau)$ is the biological factor of the emergence of new infected (general), and $I_p(t)$ is social infectiousness management – a factor that differs from country to country and varies over time. The first two days and on the last one (16th) the probability of recovery is 0:
 $b(0)=0; b(1)=0 \quad \{ b(\tau) = 0, \tau \in [0,2] \}$
 $b(16)=0.$
For definiteness, we normalize one of the factors:
$$\sum_{\tau'=0}^{16} b(\tau') = 1 \qquad \{ \int_0^{\tau_{max}} b(\tau') d\tau' = 1 \}$$

**Summing up.** The accepted hypotheses and their mathematical formalization determine the model, which for given:
 • general biological functions $b(\tau), \mu(\tau), f(\tau)$,
 • (country-specific) social administrations $I_p(t), F_p(t)$,

• a given initial distribution of the UDI $n_p(0,\tau)$

allows for each of the populations under consideration to calculate $n_p(t,\tau)$ – the temporal dynamics of the UDI distributed over the DI.

## BALANCED IDENTIFICATION

The identification problem consists of finding unknown common biological functions $b(\tau)$, $\mu(\tau)$, $f(\tau)$ and the population specific functions $F_p(t)$, $I_p(t)$, $p \in [0,P]$, $P=6$, so that the model new cases (of infected)

$$nC_p(t) = F_p(t) \sum_{\tau'=0}^{16} f(\tau') n_p(t, \tau')$$

Obviously, the statement is too vague and not correct. So we use the method of balanced identification (Sokolov, Voloshinov, 2018; Sokolov, Voloshinov, 2019). We define the procedure for selecting the entire set of unknown functions by minimizing the functional depending on the vector of regularization parameters α

$$K(b, f, \mu, I, F, n, nC, \overline{\alpha}) = \sum_{p=0}^{P} \sum_{t \in T_p} (nC_{t,p} - nC_p(t))^2 +$$

$$\alpha_1 \cdot Curv(b) + \alpha_2 \cdot Curv(f) + \alpha_3 \cdot Curv(\mu) + \sum_{p=0}^{P} \alpha_4 \cdot Curv(I_p) + \alpha_5 \cdot Curv(F_p) +$$

$$\alpha_6 \cdot \sum_{p=0}^{P} \sum_{t \in T_p} N_p(t) \to \min, \ldots$$

where $Curv(y(x)) = \int_a^b (y''(x))^2 dx$ defines the curvature of the $y(x)$ function.

The first term of the functional is a measure of the model's fit to measurements (standard deviation), addends 2-6 are regularizing additives, the measure of complexity of the model, which is the curvature of the function (the square of the second derivative) $b(\tau)$, $\mu(\tau)$, $f(\tau)$, $F_p(t)$ and $I_p(t)$. Finally, the last addend minimizes the total number of UDI, so that the solution found is as small as possible (lower bound).

The regularization parameters α are selected by minimizing the mean square error of cross-validation.

## MODELING RESULTS AND DISCUSSION

As a result of the model (7 populations) identification based on data on new cases, the general biological functions ($b(\tau)$, $\mu(\tau)$, $f(\tau)$) were determined, and for each population social controls ($I_p(t)$, $Fp(t)$), the solution itself ($n_p(t, \tau)$) and a number of additional indicators convenient for analyzing the results were found.

*Common biological functions*

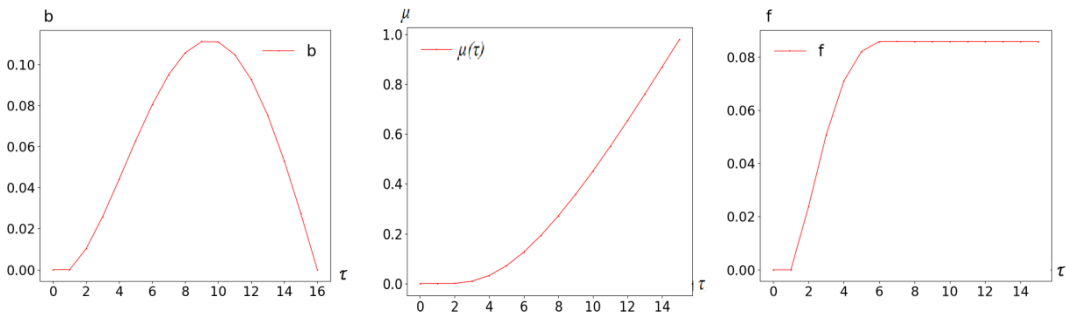

Figure 3. Biological functions of the virus-human interaction: $b(\tau)$ is the DI-dependent infectivity (contagiousness) factor, $\mu(\tau)$ is the recovery (without detection) and $f(\tau)$ is the DI-dependent detection factor, $\tau$ is the duration of infection (DI).

Figure 4 shows graphs of generally accepted and more obvious modifications (taking into account the reduction in the number of cohorts due to recovery) of biological functions of infectivity and detectability

$$b^*(\tau) = b(\tau)exp(-\int_0^\tau \mu(\tau')\,d\tau')$$

$$f^*(\tau) = f(\tau)exp(-\int_0^\tau \mu(\tau')\,d\tau')$$

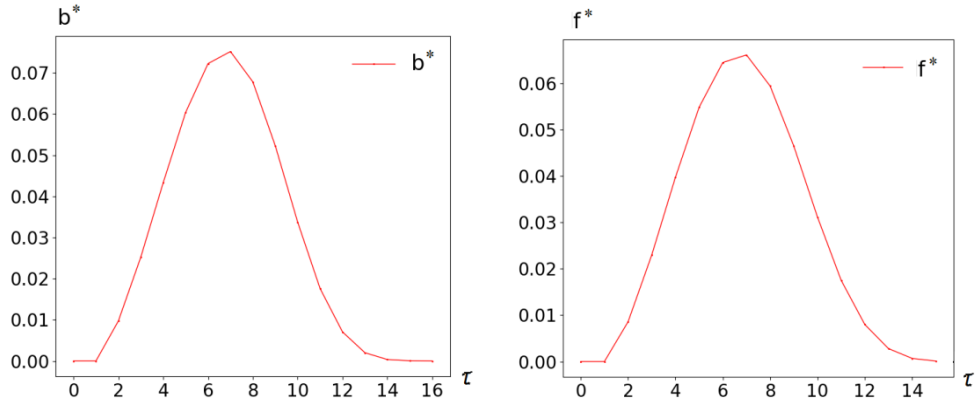

Figure 4. Modified biological functions of the interaction of the virus with humans: $b^*(\tau)$ – DI-dependent infectivity (contagiousness) factor, and $f^*(\tau)$ – DI-dependent detection factor, $\tau$ – infection duration (DI).

The maximum of infectiousness (7 days) corresponds to modern theories about COVID-19. Apparently, it is also approximately at this time that the maximum manifestation of symptoms is observed, which corresponds to the maximum of detection.

***Social management (results for different populations.*** For each population (except Russia excluding Moscow and the region), the last point with the number of infected less than 100 was taken as 0. Since different countries use different calculation methods, only a qualitative comparison of the corresponding results is possible.

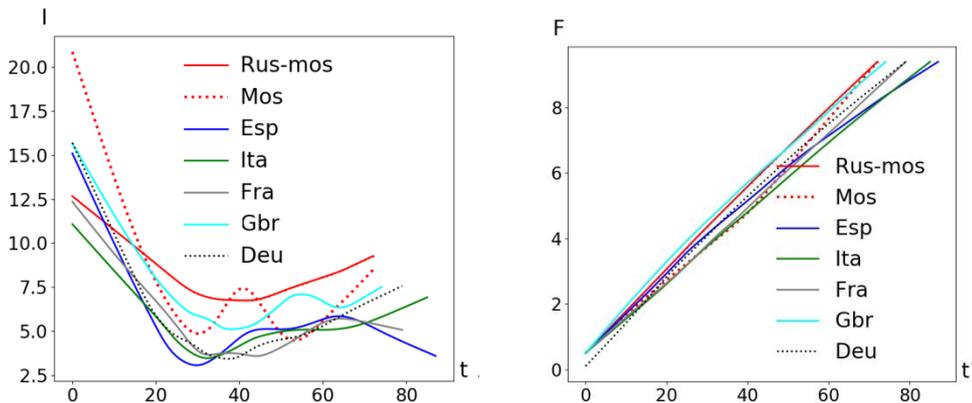

Figure 5. Social management that determines the UDI dynamics as of May 29: $I(t)$ – social management of infectiousness (restriction of contacts, self-isolation, masks, etc.); $F(t)$ – social management of detection of infected people (temperature measurement, tests, etc.)

**Additional indicates.**
*Total number of UDI*

$$N_p(t) = \sum_{\tau'=0}^{16} n_p(t,\tau') \qquad \left\{ N_p(t) = \int_0^{\tau_{max}} n(t,\tau')\, d\tau' \right\}.$$

Figure 6 shows the model curves $N_p(t)$ for all populations under study.

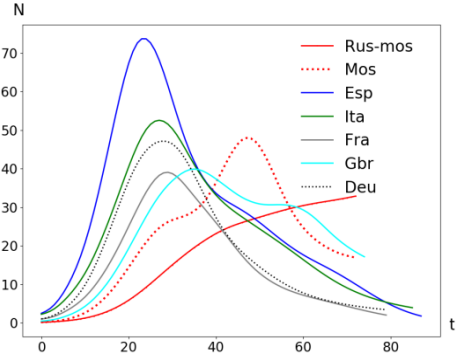

Figure 6. Dynamics of the total number of UDI – N (t) (thousand people) – lower estimate. As of May 29

*New cases (New detected cases of infection)*

$$nC_p(t) = F_p(t) \sum_{\tau'=0}^{16} f(\tau') n_p(t,\tau') \qquad \left\{ nC_p(t) = F_p(t)_p \int_0^{\tau_{max}} f(\tau') n(t,\tau')\, d\tau' \right\}.$$

It is this parameter that connects the model with the corresponding statistical indicator.

Figure 7 shows the $nC_p(t)$ model curves for all populations under study. (The curves along with the statistical data for two populations can be seen in Figure 9)

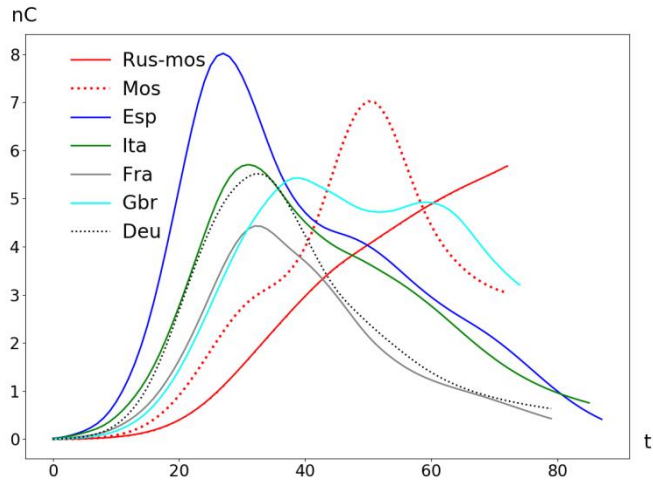

Figure 7. The dynamics of new cases $nC(t)$ (thousand people), as of May 29.

Figure 8 shows graphs of two complex indicators: the frequently used indicator R0 (reproduction index or basic reproductive number) – the average number of people infected by one patient over the entire duration of the disease – and the detection index F0 – the probability of the infected being detected during the entire duration of the disease:

$$R0_p(t) = I_p(t) \int_0^{\tau_{max}} b(\tau) \exp\left(-\int_0^{\tau} (\mu(\tau') + F_p(t) f(\tau))\, d\tau'\right) d\tau$$

$$F0_p(t) = F_p(t) \int_0^{\tau_{max}} f(\tau) exp\left(-\int_0^{\tau} (\mu(\tau') + F_p(t)f(\tau))\, d\tau'\right) d\tau$$

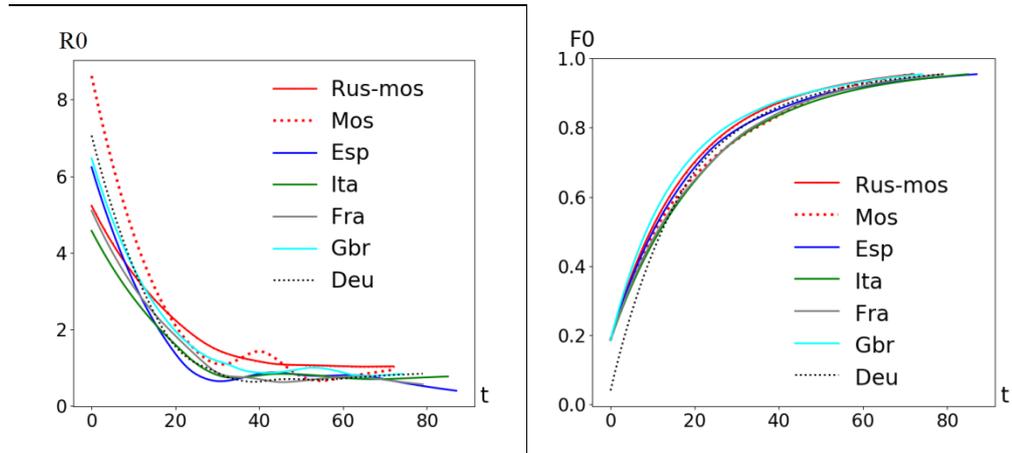

Figure 8. Reproduction index R0 (basic reproductive number) – how many people will be infected by an undetected infected during the whole time of the disease in a given setting situation (*I(t), F(t)*). Detection index F0 – probability of detecting an infected person during the whole time of the disease in a given situation (I (t), F (t)). As of May 29.

These indices (depending on *I(t), F(t)*) reflect the effectiveness of society's struggle against the epidemic. The local maximum R0 for Moscow and the region near point 40 corresponds to May 1.

### *Additional results for Russia and Moscow*
First of all, Fig. 9, shows the initial statistics for new cases of infection and the corresponding model curves.

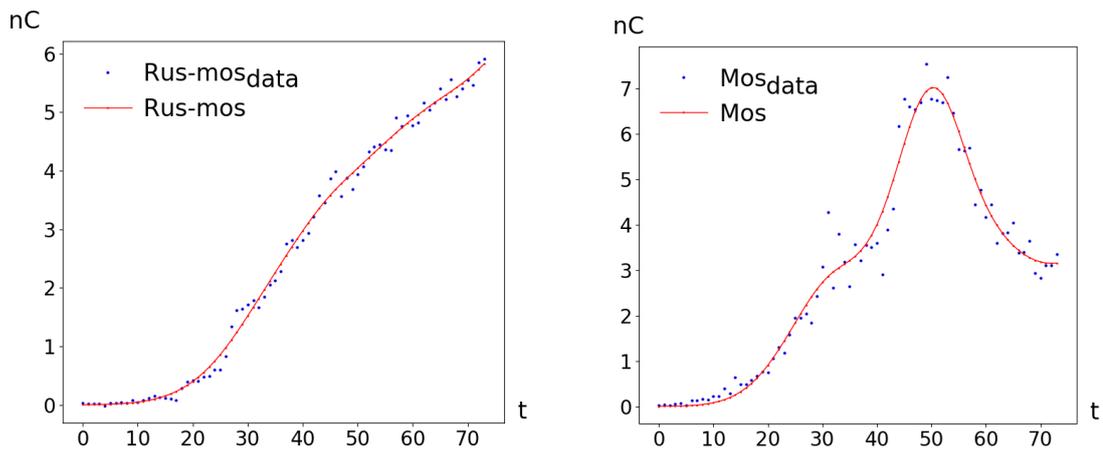

Figure 9. New cases of infection of Russia (excluding Moscow and the region) and the city of Moscow with the region. Source data *nC_t* and model curves *nC(t)* (thousand people). As of May 30.

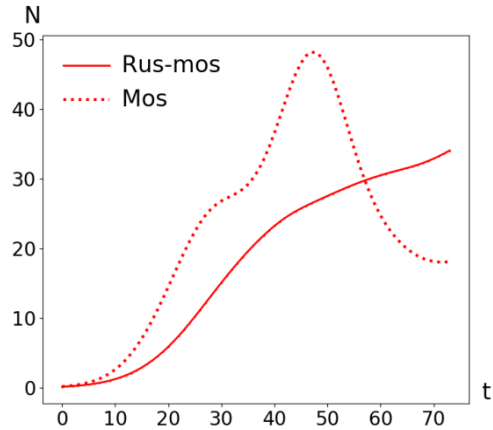

Figure 10. The dynamics of the total number of UDI for Russia without Moscow and the region and for Moscow and the region is N (t) (thousand people, lower estimate). As of May 30th. The maximum value of the UDI for Moscow and the region (about 45 thousand) corresponds to May 5.

***The total number of new UDI.*** *Addition to the UDI population.*
$$B_p(t) = n_p(t,0)$$

***The total amount of UDI decline.*** A decrease in the population of UDI occurs due to recovery and detection

$$D_p(t) = \sum_{\tau'=0}^{16} \mu(\tau')n_p(t,\tau') + F_p(t)\sum_{\tau'=0}^{16} f(\tau')n_p(t,\tau')$$

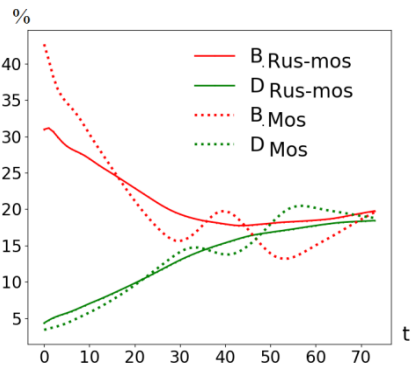

Figure 11. The increase (B) – the total number of new UDI and the decrease (D) – the total amount of UDI decline as a percentage of the total number of the UDI. The difference gives an increase in percent. The intersection points are zero growth. As of May 30.

The local maximum of the increase of the UDI population in Moscow and the region near point 40 corresponds to May 1. A similar maximum is observed in Fig. 7.

All the estimates above are lower estimates. This is due to the fact that the number of undetected infected people is not measured, but is the product of the identification problem solution, and the optimization criterion for the choice of the solution (the last summand) lowers the average number of UDI.

Additional information, such as antibody screening results, can be used to estimate the actual amount of UDI. Assuming that 14% of the Moscow population have antibodies (as per screening results), i.e. had been or are UDI and formalizing this assumption in the form of an additional identification condition, the number of UDI will increase tenfold (see Fig. 12), and the maximum value of UDI will reach 450 thousand (May 5). Under this assessment, by May 5 every 25th Moscow resident was undetected infected, and by May 30 – every 50th.

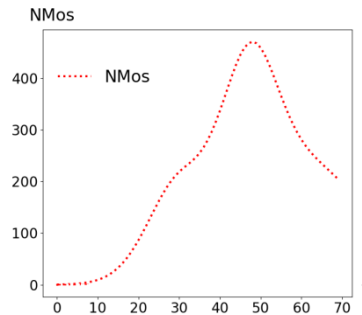

Figure 12. The dynamics of the total number of UDI in Moscow and the region - N (t) (thousand people) – an estimate corresponding to a screening of antibody carriers of 14%. As of May 28.

## CONCLUSION

The combination of knowledge about the interaction of the virus with human and human with society, with experimental data based on the balanced identification method, made it possible to choose a model with complexity corresponding to the volume and quality of the experimental material. The resulting functions are consistent with current ideas on the processes that determine the dynamics of the epidemic.

The authors did not set themselves the task of predicting the dynamics of the epidemic (such task involves a separate study, with another modified identification procedure). However, the obtained results suggest the possibility of a forecast for 5-7 days, provided that (for this period) *I(t)* and *F(t)* trends are preserved. Moreover, in this case, it seems possible to estimate the error of such prediction.

Comparison of possible models (and the choice of a model that is in good agreement with available observations) was carried out in a very short time (essentially, in a few days). This was made possible through the use of balanced identification (SvF) technology (Sokolov, Voloshinov, 2019)[1], and the SSOP[2] service designed to solve sets of independent optimization problems on computing resources connected to the optimization modeling subsystem https://optmod.distcomp.org, deployed on the platform Everest (Sukhoroslov O. et al., 2015).

The technology of balanced identification supports the (evolutionary) modification of the model – a new (usually more complex) model is built on the basis of the previous one, while the previously found solution is used as an initial approximation in finding a solution for a new model. For example, when new data or knowledge appears, it seems justified to change the accepted hypotheses, primarily taking immunity into account and dividing the population into additional groups. This approach has already been used to modify the model to take screening data into account (see Fig. 12).

*This research:*
- *has been carried out using computing resources of the federal collective usage center Complex for Simulation and Data Processing for Mega-science Facilities at NRC "Kurchatov Institute", http://ckp.nrcki.ru/.*
- *was supported in part through computational resources of HPC facilities at NRU HSE ".*
- *was financially supported by the Russian Foundation for Basic Research grant (projects 20-07-00701 and 18-07-01269).*

---

[1] https://gitlab.com/sashasok/svf

[2] https://optmod.distcomp.org/apps/vladimirv/solve-set-opt-probs

# APPENDIX

A fragment of the task file with a formal description of the model for calculations (balanced identification technology).

```
tmin = 0
maxF = 200
maxFf = 0.6
R0 = 170
RECpart = 0.1
tam = 16
st = 1
GRID:    ta  = [ 0, tam,    st, tai ]
         ta1 = [ 0, tam-st, st, tai ]
Var:     b ( ta ) >= 0; b(0)=0; b(st)=0; b(tam)=0;  ∫r(ta,dta*b(ta)) = 1
         f ( ta1 ) >= 0; f(0) = 0; f(1) = 0;  ∫r(ta1,dta1*f(ta1)) = 1
         miu( ta1 ) >= 0; <=1; miu(tam-st)=.98; miu(0)=0; miu(st)=0;
         l ( ta ) >= 0; l(0)=1; l(ta1+st) = l(ta1)* ( 1-miu(ta1) )
         bl ( ta ) >= 0; bl = b * l
         R0mult   >= 0; R0mult = ∫( 0,tam, dta*bl(ta) )
Param:   l0(ta) = l0(ta).sol
Select   iso_code, date, new_cases As nC4, ROWNUM AS t4, ROWNUM AS t4p \
         from ../owid-covid-data5-10.xlsx where iso_code==4 and date >=20200223  # ITA
GRID:    t4  = [ tmin   , , st, ti  ]     # Time - number of point (first=0)
         t4p = [ tmin+st, , st, ti  ]
MakeSets_byParam t4 7 #1                            # CV procedure parameters

    Var:         nC4 ( t4 )  >= 0
         n4 (t4, ta)>= 0;
         N4  ( t4 ) >= 0;  N4 = ∫r(ta,dta*n4(t4,ta)) ;
         n04 >= 0;
         np4 >= 0;
          I4  ( t4 ) >= 0;        ∫r(0,t4.max,dt4*I4(t4)) = t4.max+1;
          II4  >= 0; II4 * I4 <= R0
         F4  ( t4 )  >= 0; d/dt4(F4) >= 0;    ∫r(0,t4.max,dt4*F4(t4)) = t4.max+1;
         FF4    >= 0; FF4 * F4 >= 0.01;  FF4 * F4 * f <= maxFf; FF4 * F4<=maxF
         D4 (t4) >=0;  D4(t4)=∫r(ta1,dta1*(miu(ta1)+FF4*F4(t4)*f(ta1)-miu(ta1) \
                *FF4*F4(t4)*f(ta1) )*n4(t4,ta1) )
          B4  ( t4 )  >= 0;
EQ:      if ti+st <= t4.max:  (n4(t4+st,ta1+st)-n4(t4,ta1))/st = -( miu(ta1)+FF4*F4(t4)*f(ta1)-\
                              miu(ta1)*FF4*F4(t4)*f(ta1) )*n4(t4,ta1)
         B4 (t4) = II4 * I4(t4) * ∫r(ta,dta*b(ta)*n4(t4,ta))
         n4(t4p,0) = B4(t4p-st)
         n4(tmin,ta) = n04 * l0(ta) *(np4**ta)
         nC4(t4)  = ∫r(ta1,dta1*( FF4*F4(t4)*f(ta1)-miu(ta1)*FF4*F4(t4)*f(ta1) )*n4(t4,ta1) )
         ∫r(ta1,dta1*miu(ta1)*n4(t4,ta1)) >= RECpart * N4(t4)
```

1212